\documentclass[article]{emulateapj}

\usepackage{natbib}
\usepackage{graphics}
\usepackage{threeparttable}
\usepackage{amsmath}
\usepackage{bm}

\shorttitle{DENSE MOLECULAR GAS IN THE CND OF NGC 1068}
\shortauthors{KAMENETZKY ET AL.}

\newcommand{\unit}[1]{\ensuremath{\, \mathrm{#1}}}
\newcommand{\ms}{M$_{\odot}$}
\newcommand{\ls}{L$_{\odot}$}
\newcommand{\kmspc}{$\,\rm km\,s^{-1}\,pc^{-1}$}
\newcommand{\kms}{$\,\rm km\,s^{-1}$}
\newcommand{\kkms}{$\,\mathrm{K\,km\,s^{-1}}$}

\newcommand{\hco}{HCO$^{+}$}
\newcommand{\as}{$^{\prime\prime}$}

\newcommand{\ttt}{$\tau_{\rm 225 GHz}$}

\newcommand{\eqq}{\!=\!}  
\newcommand{\too}{\!\rightarrow\!} 
\newcommand{\jone}{{$\rm J\eqq 1\too0$}}
\newcommand{\jtwo}{{$ \rm J\eqq2\too1$}}
\newcommand{\jthree}{{$\rm J\eqq3\too2$}}
\newcommand{\jfour}{{$\rm J\eqq4\too3$}}
\newcommand{\jfive}{{$\rm J\eqq5\too4$}}
\newcommand{\jsix}{{$\rm J\eqq6\too5$}}
\newcommand{\jseven}{{$\rm J\eqq7\too6$}}
\newcommand{\jthirteen}{{$\rm J\eqq13\too12$}}
\newcommand{\jseventeen}{{$\rm J\eqq17\too16$}}

\begin{document}
\bibliographystyle{apj_w_etal}
\title{The Dense Molecular Gas in the Circumnuclear Disk of NGC 1068}

\author{J.~Kamenetzky\altaffilmark{1},
  J.~Glenn\altaffilmark{1},
  P.R.~Maloney\altaffilmark{1}
  J.E.~Aguirre\altaffilmark{2},
  J.J.~Bock\altaffilmark{3},
  C.M.~Bradford\altaffilmark{3},
  L.~Earle\altaffilmark{1,4},
  H.~Inami\altaffilmark{5},
  H.~Matsuhara\altaffilmark{5},
  E.J.~Murphy\altaffilmark{6},
  B.J.~Naylor\altaffilmark{3},
  H.T.~Nguyen\altaffilmark{3},
  J.~Zmuidzinas\altaffilmark{3}}

\altaffiltext{1}{Center for Astrophysics and Space Astronomy, 389-UCB, 
University of Colorado, Boulder, CO, 80303} 

\altaffiltext{2}{Department of Physics, University of Pennsylvania, Philadelphia, PA, 19104}

\altaffiltext{3}{NASA Jet Propulsion Laboratory, Pasadena, CA, 91109} 

\altaffiltext{4}{National Renewable Energy Laboratory, Golden, CO, 80401}

\altaffiltext{5}{Institute for Space and Astronautical Science, Japan
Aerospace and Exploration Agency, Sagamihara, Japan}

\altaffiltext{6}{Infrared Processing and Analysis Center, Pasadena, CA, 91125}

\begin{abstract}

We present a 190-307 GHz broadband spectrum obtained with Z-Spec of NGC 1068 with new measurements of molecular rotational transitions.  After combining our measurements with those previously published and considering the specific geometry of this Seyfert 2 galaxy, we conduct a multi-species Bayesian likelihood analysis of the density, temperature, and relative molecular abundances of HCN, HNC, CS, and \hco.  We find that these molecules trace warm (T $>100$ K) gas of H$_2$ number densities $10^{4.2} - 10^{4.9}$ cm$^{-3}$.  Our models also place strong constraints on the column densities and relative abundances of these molecules, as well as on the total mass in the circumnuclear disk.  Using the uniform calibration afforded by the broad Z-Spec bandpass, we compare our line ratios to X-ray dominated region (XDR) and photon-dominated region models.  The majority of our line ratios are consistent with the XDR models at the densities indicated by the likelihood analysis, lending substantial support to the emerging interpretation that the energetics in the circumnuclear disk of NGC 1068 are dominated by accretion onto an active galactic nucleus.

\end{abstract}

\subjectheadings{Galaxies: Individual: NGC 1068 - Galaxies: Seyfert - Galaxies: ISM - ISM: molecules}


\section{Introduction}\label{sec:intro}

Molecular transitions in the millimeter and submillimeter bands are well suited to characterize the physical conditions of dense gas in active galactic nuclei (AGN) or starburst (SB) regions.  Starburst regions are associated with far-UV emission from O and B stars and corresponding photon-dominated regions (PDRs, \citet{Tielens:1985}, which characterize cloud surfaces), while accretion onto an AGN creates an X-ray dominated region (XDR, \citet{Maloney:1996}, which characterize cloud volumes).  These two different types of regions can be distinguished by the signatures of the molecular gas chemistry; making such a distinction has been the focus of recent research in many (ultra-)luminous infrared galaxies or (U)LIRGs.  However, we often must understand the physical environment of the dense gas before we can properly diagnose the XDR or PDR nature of it.  For example, the predicted line ratios for XDR or PDR scenarios vary with density.  Therefore it is important to understand the physical conditions of the gas before using (sub)millimeter lines to provide key information about the dominant energy source heating the gas.

NGC 1068 (M77) is often cited as the prototypical Seyfert 2 galaxy; it is a barred spiral hosting an AGN that has been extensively studied at many wavelengths.  At a redshift of 1137 \kms\ \citep{Huchra:1999}, 1\as\ $\approx$ 78 pc assuming H$_0 = 70$ \kms\ Mpc$^{-1}$.  With an infrared luminosity of $2 \times 10^{11}$ \ls, it is classified as a LIRG \citep{Sanders:2003}.  The major features of NGC 1068 are
the large star-forming arms, a central circumnuclear disk (CND), and a jet
emerging in the northeast direction from the CND.  The star-forming arms,
seen most prominently in CO line emission, essentially form a large ring of
30\as\ diameter (about 2.3 kpc, Figure 2 of \citet{Helfer:1995}, Figure 1 of
\citet{Schinnerer:2000}).  The CND is approximately 4\as\ diameter (312 pc), from
which arises emission of molecular high-density gas tracers (e.g., CS, \hco, HCN, HNC), as can be seen in many maps (such as
\citet{Helfer:1995,GarciaBurillo:2008,Krips:2008,Bayet:2009} and the 1 mm
and 3 mm continuum maps of \citet{Krips:2006}).  The CND itself can be
resolved into two knots (East and West) with different velocity profiles
\citep{Usero:2004}.  The geometry of NGC 1068 is a significant consideration
for single-dish observations, because the two very different regions (the
CND and SF ring) are blended by most (sub)millimeter telescope beams.  For example, for our $\sim$ 30\as\ beam, we are measuring some emission from the SF ring on the outer edges of our beam, and also emission from the central CND concentrated in the middle.  In the case of CO (and the continuum) we are measuring both components.  For the high-density tracing molecules, however, the emission is concentrated only in the center of our beam (from the CND), as can be seen in the aforementioned interferometric maps.  The physical area that we are modeling for this paper is the CND only; we address how we correct for possible contamination from the starburst ring in \S \ref{sec:fsb}.

The black hole mass at the center of the CND, as inferred from water maser emission, is $1.7 \times 10^7$ \ms\ \citep{Greenhill:1997}.  The bolometric luminosity of the entire galaxy (ring + CND, $2.5-3 \times 10^{11}$ \ls) is determined mainly by the mid-IR; about half of this emission comes from the starforming ring, so the bolometric luminosity of the central 4\as\ in diameter has been estimated as $1.5 \times 10^{11}$ \ls \citep{Bock:2000}.  Though this is the luminosity visible to us from the CND, the AGN itself has a higher intrinsic luminosity that is well shielded by its obscuring torus.  Using [OIV] 26 $\mu$m as a calibrator, \citet{Rigby:2009} infer an intrinsic (but highly obscured) AGN luminosity of $(3.1 \pm 0.9) \times 10^{11}$ \ls, consistent with the estimation of \citet{Bock:2000} ($3 \times 10^{11}$ \ls) based on L$_{bol} \sim 30$ L$_{2-10keV}$.  The high amount of the AGN's luminosity that is absorbed  indicates that the AGN can be extremely important in powering the CND emission.

Several recent studies indicate that the CND of NGC 1068 is an XDR \citep{Usero:2004,Perez:2009,Krips:2008,GarciaBurillo:2010}.  There are a variety of diagnostics of XDR/PDR environments that can be utilized with millimeter-range molecular transitions.  AGN tend to have lower intensity ratios of \hco/HCN (\jone) and higher ratios of HCN/CO (\jone, \citet{Krips:2008}).  \hco\ [(\jthree)/(\jone)], HCN [(\jtwo)/(\jone)], and HCN [(\jthree)/(\jone)] intensity ratios also appear lower in AGN than in SB.  In Krips' study of 12 nearby galaxies with varying degrees of AGN/starburst activity, NGC 1068 was the most characteristic of AGN line ratios (and generally appearing completely opposite from M82, a prototypical starburst galaxy).  Furthermore, it is predicted that CN is more abundant than HCN in XDR scenarios due to dissociation (HCN $\rightarrow$ H + CN, but then CN is particularly robust to further dissociation, \citet{Lepp:1996p3242,Meijerink:2007}).

By observing NGC 1068 with Z-Spec, a broadband millimeter-wave spectrometer, we are able to measure the continuum simultaneously with multiple molecular transitions.  We present observations of HCN, \hco, and HNC which complement the current literature as well as new transitions of CS.  The paper is organized as follows.  The Z-Spec instrument, our observing procedure, and the spectrum are described in \S \ref{sec:obs}.  We use the measurements in our spectrum in conjunction with data from the literature to  model first the physical conditions of the molecular gas in the CND of NGC 1068 using RADEX \citep{vanderTak:2007}.  The modeling procedure is described in \S \ref{sec:analysis}, with discussion of the results beginning in \S \ref{sec:disc}.  We next use the derived molecular abundances and the line ratios within Z-Spec's band to demonstrate the XDR nature of the CND in \S \ref{sec:diagnostics}, followed by conclusions in \S\ref{sec:concl}.

\section{Observations with Z-Spec}\label{sec:obs}

In this section, we first describe the Z-Spec instrument and observations of NGC 1068.  We then present our spectrum along with a description of our spectral fitting procedure and a short discussion of the fit results.

\begin{figure*}
\centering
\scalebox{0.7}{\includegraphics*{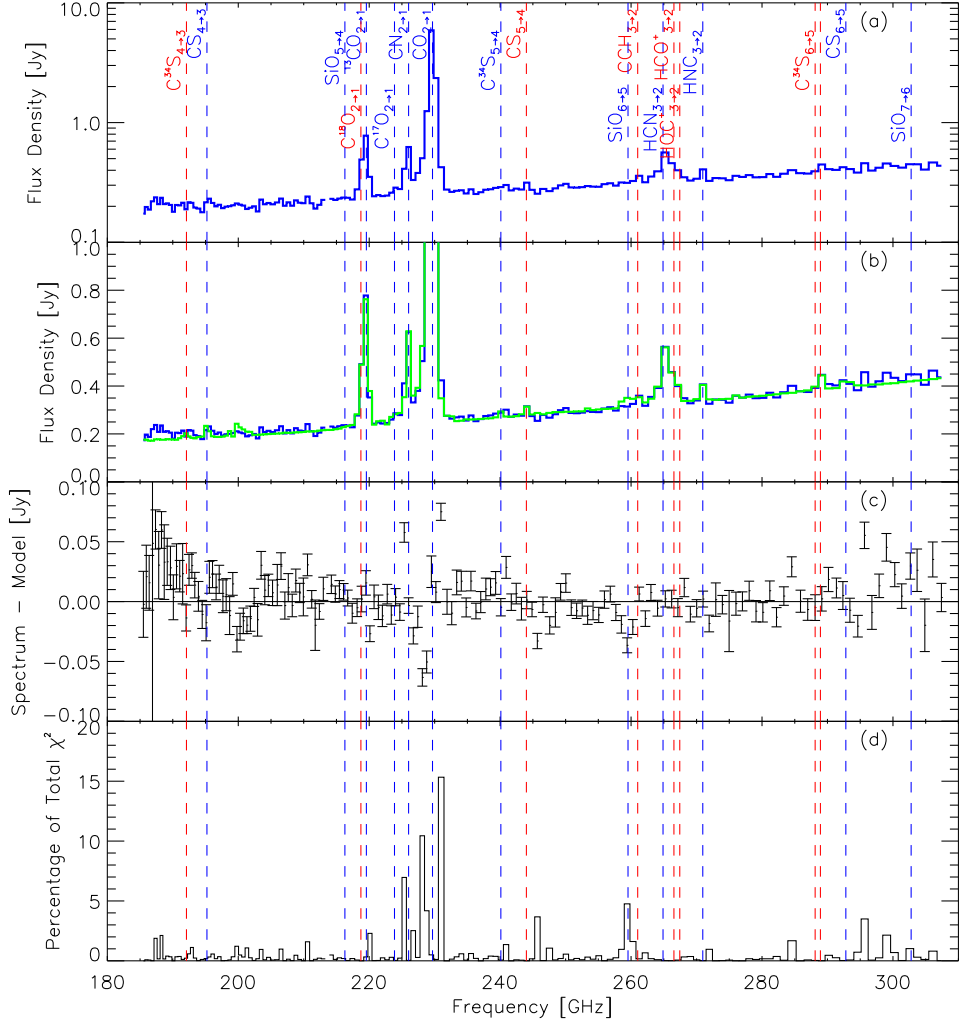}}
\caption[fig:spectrum]{Spectrum of NGC 1068 observed by Z-Spec on the CSO, shown in logarithmic (a) and linear (b) scale.  The spectral fit (described in \S \ref{sec:fitting}) is overplotted in green on the linear plot.  Vertical lines indicate the frequencies of known molecular transitions of importance, including the prominent CO \jtwo\ line (vertical lines alternate between red and blue for clarity only).  Transitions of SiO are not included in the fit but are shown here for reference.  The results for HOC$^+$ \jthree\ and C$^{34}$S \jfour\ are only upper limits.  The unlabeled line at $\sim$ 290 GHz is an unidentified feature that is modeled as a Gaussian in the fit.  Features above CS \jsix\ are not included due to the high channel-to-channel variation at the end of the band due to atmospheric noise.  The features in the spectral fit at 200 and 260 GHz are instrumental artifacts described at the end of \S \ref{sec:fitting}.  Panel (c) illustrates the difference (residual) between our model fit and the spectrum, with uncertainties.  Panel (d) shows the contribution from each channel to the total $\chi^2$; the ten channels around the CO \jtwo\ line are responsible for 40\% of the total $\chi^2$ of 5.1.}\label{fig:spectrum}
\end{figure*}

\subsection{The Z-Spec Instrument}\label{sec:zspec}

Z-Spec is a broadband (190-307 GHz) millimeter-wave grating spectrometer which we have used at the Caltech Submillimeter Observatory (CSO).  Its large bandwidth allows 
simultaneous observations of multiple molecular rotational transitions along
with the underlying continuum.  The detector array is composed of 160
silicon-nitride micromesh bolometers cooled to 60 mK by an adiabatic
demagnetization refrigerator (ADR) and a closed-cycle $^3$He refrigerator.
Z-Spec's compact design is acheived via a WaFIRS (Waveguide Far IR
Spectrometer) design utilizing a parallel-plate waveguide and curved Rowland diffraction grating \citep{Bradford:2003}.  Z-Spec's spectral resolution is approximately 900 MHz
at the band center, but varies from 500 MHz (700 \kms) at the low frequency end to 1300 MHz (1200 \kms) at the high frequency end of the band.  To minimize susceptibility to  atmospheric fluctuations and to subtract sky background emission, we use a chop-and-nod mode.  For NGC 1068 observations, our nod interval was 20 seconds, and the chop frequency was 0.95 Hz with a throw of 90\as.  More details about the instrument can be found in \citet{Naylor:2003}, \citet{Bradford:2004}, \citet{Earle:2006}, and \citet{Inami:2008}.

We observed NGC 1068 over three nights in late January 2007 for a total of
4.21 hours of integration time (including chopping/nodding, meaning one half of this time was on source).  On January
24 and 28, the optical depth at 225 GHz, \ttt\, was steadily $\sim$ 0.05, while on January 27 it was slightly worse at \ttt\ $ \sim$ 0.070-0.086.  The median sensitivity (channel uncertainty times square root of integration time) was 1.13 Jy s$^{1/2}$, some two to three times greater than we typically achieve in observations of faint sources.  We attribute the excess to systematic effects due to the bright source.

Z-Spec's calibration relies on using a set of planetary observations with varying observing conditions (bolometer loading and bath temperature) and fitting the dependence of each individual bolometer's response on its operating voltage, as described in \citet{Bradford:2009}.  Because our observations of NGC 1068 were conducted before the calibration curves described in the aforementioned paper, it was more appropriate to calibrate these observations with local-time observations of Uranus.  The total integrated continuum flux measured with the Bradford and Uranus calibrations agree to within 5\%.

\subsection{Continuum Analysis and Line Fitting Procedure}\label{sec:fitting}

\begin{deluxetable}{ c c c c c}
   \tablecaption{Z-Spec Detections in NGC1068\label{table:fit}}
   \tabletypesize{\footnotesize}
   \tablewidth{0pt}
   \tablehead{
   \colhead{Transition} & \colhead{$\nu_{rest}$} & \colhead{$\int F_\nu dv$} & \colhead{$\int T_A dv$} & \colhead{S/N\tablenotemark{a}}\\
   \colhead{}        & \colhead{[GHz]}        & \colhead{[Jy \kms]}       & \colhead{[K \kms]}      & \colhead{}}
   \startdata
CO$_{2 \rightarrow 1}$ & 230.54 & 8366 & 183.7 & 436\\
$^{13}$CO$_{2 \rightarrow 1}$ & 220.4 & 712 & 15.8 & 54.4\\
C$^{17}$O$_{2 \rightarrow 1}$ & 224.7 & 31 & 0.7 & 3.0\\
C$^{18}$O$_{2 \rightarrow 1}$ & 219.6 & 236 & 5.2 & 16.6\\
CCH$_{3 \rightarrow 2}$ & 262.25 & 57 & 1.2 & 6.1\\
CN$_{2 \rightarrow 1}$ & 226.87 & 496 & 10.9 & 54.2\\
CS$_{4 \rightarrow 3}$ & 195.95 & 68 & 1.6 & 3.6\\
CS$_{5 \rightarrow 4}$ & 244.94 & 58 & 1.2 & 5.4\\
CS$_{6 \rightarrow 5}$ & 293.91  & 58 & 1.2 & 3.0\\
C$^{34}$S$_{5 \rightarrow 4}$ & 241.02 & 40 & 0.9 & 3.8\\
HCN$_{3 \rightarrow 2}$ & 265.89  & 406 & 8.5 & 33.1\\
HCO$^+$$_{3 \rightarrow 2}$ & 267.56 & 267 & 5.6 & 15.9\\
HNC$_{3 \rightarrow 2}$ & 271.98  & 104 & 2.2 & 10.3\\
Unidentified\tablenotemark{b} & 290.02 & 93 & 1.9 & 6.0\\
C$^{34}$S$_{4 \rightarrow 3}$\tablenotemark{c} & 192.82 & 32 & 0.7 & 2.2\\
HOC$^+$$_{3 \rightarrow 2}$\tablenotemark{c} & 268.45 & 19 & 0.4 & 1.2
   \enddata
	\tablenotetext{a}{S/N estimates do not include calibration error, which is estimated to be 10\%.}
	\tablenotetext{b}{The unidentified feature at 290 GHz is described in \S \ref{sec:fitting}.}
   \tablenotetext{c}{These two entries should be regarded as only upper limits.}  
\end{deluxetable}

The Z-Spec spectrum of NGC 1068 is shown in Figure \ref{fig:spectrum}.  Our line-fitting procedure treats each emission line as a single-component Gaussian.  Each channel's spectral response profile has been previously measured and is incorporated into the line-fitting routine.  Higher-resolution spectra reveal more detailed structure, such as two or more components for the high-density tracers and generally three components for CO (e.g. \citet{Perez:2007,Krips:2008}), but we cannot resolve these structures.  Because Z-Spec cannot resolve line widths below $\sim$ 1000 \kms, the line width was set to 240 \kms\ based on the approximate CO \jtwo\ linewidth in \citet{Israel:2009} (some measured NGC 1068 linewidths have been higher or lower).  One channel (at 213.3 GHz) was not operating during our observations and is excluded from our fit.  In our line fitting procedure, we choose the lines to be fit based on a list of the expected transitions in our band, and those which are not reliably detected are excluded and then the spectrum refit.  Rest line frequencies are taken from the online Leiden Atomic and Molecular Database \citep{Schoier:2005}.

\citet{Krips:2006} measured the continuum emission from the core and jet of NGC 1068; at 230 GHz, the measured continuum flux from this region is only $\sim 10\%$ of our measured continuum.  Therefore, the continuum measured by our beam is dominated by the starburst ring, but a small contribution is also present from the central disk and jet.  We add their ``core + jet" model (their Figure 3, top panel) to the (beam-scaled) 34 K greybody of the starburst ring measured by \citet{Spinoglio:2005}, so that our continuum fit becomes:

\begin{equation}
\begin{split}
F_{\nu} =& A \bigg( \frac{\nu}{240 \rm GHz} \bigg)^{B-2} \Omega B_{\nu}(T)\bigg(1-e^{-(\nu/\nu_0)^{\beta}} \bigg)\\
         & + F_{0,core+jet} \bigg(\frac{\nu}{230 \rm GHz} \bigg)^{-\alpha} \textup{Jy}.
\end{split}
\end{equation}

\noindent The first term is the beam-scaled greybody, and $\Omega$ is the source size in steradians; we estimate $\Omega$ to be approximately $1.66 \times 10^{-8}$ sr, based on the 30\as\ diameter of the starburst ring.  Since the emission from the ring actually only occupies a portion of this total area, this is likely an overestimation that can be accounted for with the free parameter A.  $B_\nu(T)$ is a blackbody of dust temperature $T = 34$ K, and we take $\nu_0 = 3000$ GHz (which corresponds to $\lambda_0 = 100 \ \mu$m) and $\beta = 2$ for the dust emissivity spectral index \citep{Spinoglio:2005}.  The second term is the power-law contribution from the disk and jet, where $F_{0,core+jet}$ is the measured flux at 230 GHz and $\alpha$ is the measured power-law scaling.  We use $F_{0,core+jet} = 0.028$ Jy and $\alpha = 0.9$ from \citet{Krips:2006}.

A and B are the free parameters in our fit; B is meant to represent the beam-scaling, because our beamsize changes with frequency and thus our coupling with the starburst ring will also change.  B would be expected to be 0 for a beam-filling source and 2 for a point-source.  We note that at these frequencies, the beam-scaled greybody (first term) approximates to a powerlaw, where

\begin{equation}
F_{\nu} \propto A \Omega T \nu_0^{-\beta} \ \nu^{B+\beta}.
\end{equation}

\noindent This means that our best-fit A value is really degenerate with $\Omega$ and $T$, and B is degenerate with $\beta$.  Our best-fit model yields $A = 0.0446 \pm 0.0002$ and $B = 0.421 \pm 0.027$.  Our goal with this model is to determine the best-fit continuum in order to measure the line fluxes accurately, not necessarily to properly model the physical conditions of the continuum.  Therefore, our best-fit parameters may be degenerate with uncertainties in other physical parameters as described above.  For example, if we were to assume $\beta < 2$, the difference would be made up in the parameter B.  A is much less than 1 likely because the dust continuum emission does not entirely fill the beam (geometrically), and the gas emitting the continuum likely has some filling factor less than 1.  We would still expect $B > 0$ because the changing size of the beam does encompass a varying amount of the starburst ring with frequency (i.e. we expect a frequency dependence for the relative fraction of coupling to the ring).  Finally, we note that the core/jet emission makes only a small contribution to the overall continuum, but it does affect the shape of the continuum and especially improves our fit at the higher end of our band; without it, we would likely overestimate the CS \jsix\ flux by over 10\%.  In summary, this model adequately fits our continuum spectrum with reasonable physical parameters.

The resulting measurements are given in Table \ref{table:fit}.  This fit had a reduced $\chi^2$ of 5.1 with 140 degrees of freedom.  As can be seen in the bottom panel of Figure \ref{fig:spectrum}, a large percentage of the total $\chi^2$ comes from the CO line, which is not well fit by a single-component Gaussian.  We do not have the resolution to be able to better model this line, whether its assymetry is due to kinematic structure or blending of other lines.  If we exclude the bins containing the CO line, the total $\chi^2$ is $<$ 3.  However, we do not use CO in our radiative transfer analysis, so its poor fit is not problematic for the rest of our results.  

Our fit includes one unidentified spectral feature at a rest frequency of approximately 290 GHz.  This frequency may correspond to CH$_3$CCH \jseventeen\ or H$_2$CO $\rm J\eqq4_{04}\too3_{03}$, but we cannot identify other features of these molecules that might be in our band.  Likely, this feature is a combination of emission features that we cannot resolve, and is therefore left as unidentified.  Though other features may be present at the upper end of the spectrum, the channel-to-channel variation increases significantly above the CS \jsix\ transition due to imperfect atmospheric subtraction, so we do not attempt to define spectral features above this frequency.  SiO \jseven\ could possibly be identified in this range, but because we cannot identify the two lower-J transitions in our band, we do not fit this feature either.  Below this frequency range, there are a few other features of note that are not fitted.  The first is a possible 3-sigma feature at approximately 284 GHz; given our inability to know the exact line center and no particularly strong transition expected that at wavelength (though there are a few transitions of methanol, CH$_3$OH, around that frequency), we do not fit this line.  Though the same difficulties apply to the unidentified line at 290 GHz, its significance was twice that of this line, and warranted inclusion in the fit.  The apparent features at 200 and 260 GHz are not lines, but part of the continuum model, having been passed through our measured line profiles which included the instrumental sidelobes of the CO \jtwo\ transition.  Future use of Z-Spec will include modification of the line profiles to exclude these artifacts.

\subsection{Spectrum Results}\label{sec:results}

When comparing our fluxes to those previously published, care must be taken to account for different beam sizes because of beam dilution.  Throughout this paper, we correct for beam dilution by dividing all Rayleigh-Jeans velocity integrated temperatures by

\begin{equation}f = \frac{\theta_{source}^2}{\theta_{source}^2 + \theta_{beam}^2}, \end{equation}\label{eqn:sizecorr}

\noindent where $\theta_{beam}$ is the full-width at half maximum of the beam profile and $\theta_{source}$ is the FWHM of the source.  For all high density tracers, we assume a source size of 4\as\ (see discussion on the geometry of NGC 1068 in \S \ref{sec:intro}).  Others have been assuming smaller source sizes such as 1.5\as\ \citep{Perez:2009}, but we have found that a smaller source size does not significantly change our results for the physical conditions of the CND.

For the rest of this paper, we will be focusing on high-density tracer emission from the CND, but first we find it appropriate to comment on the most prominent feature of our spectrum, $^{12}$CO \jtwo, mostly from the starburst ring.  A source size of 30\as\ has been estimated from maps of CO emission (representing the diameter of the starburst ring, see \S \ref{sec:intro}).  After correction for source size and comparing to the one reported \jtwo\ transition from \citet{Perez:2007} and three reported \jtwo\ transitions from \citet{Israel:2009}, Z-Spec's measured flux is slightly higher than all four others.  However, after adding in a 15\% calibration uncertainty to each measurement, they agree within the uncertainties.  This gives us an indication that our calibration is consistent with others.

The transitions useful to the study of the dense nuclear core of NGC 1068, the subject of the remainder of this paper, are those of HCO$^+$, HCN, HNC, CN, and CS.  The CS \jfour\ and CS \jsix\ measurements are new, and when combined with previous CS \jthree, CS \jfive, and CS \jseven\ measurements, provide a  well-sampled ladder for radiative transfer modeling.  We do not include CN in our models because we cannot accurately separate its hyperfine transitions; our reported CN intensity includes all 18 hyperfine lines included in CN \jtwo.


\section{Modeling of Molecular Gas Physical Conditions}\label{sec:analysis}

We sought to estimate the most probable physical conditions of NGC 1068's dense core by simultaneously modeling a number of molecular lines.  Rather than just find the single most probable set of conditions, we examined a large parameter space and determined the most probable set of conditions by marginalizing over all possibilites.  To do so, we used Bayes' theorem to describe the likelihood of a particular set of model parameters given our observational measurements, as described in \citet{Ward:2003}.  We have extended this analysis to examine multiple molecules at once, as in \citet{Naylor:2010}.  Here, we first describe RADEX, which is used to create a grid of expected line fluxes for a range of physical parameters, then we detail the likelihood analysis and the specific considerations for NGC 1068.

\subsection{RADEX}\label{sec:RADEX}

To investigate the physical parameters, such as temperature and density, that produced our measured line fluxes, we used RADEX, a non-LTE code freely available from the Leiden Atomic and Molecular Database \citep{vanderTak:2007}.  We used the large velocity gradient (LVG) model to perform statistical equilibrium calculations using the escape probability in an expanding spherical cloud, but the results are insensitive to the precise form of the escape probability.  When provided with the proper input data, RADEX calculates the level populations in the optically thin limit considering the background radiation field and then calculates the optical depths for the lines.  The code continues calculating new level populations using new optical depth values until the two converge on a consistent solution.  The program then outputs the resulting line intensities as background-subtracted Rayleigh-Jeans equivalent radiation temperatures.  

The background radiation field can be as simple as a 2.73 K blackbody to represent the cosmic microwave background (CMB).  However, the continuum emission from the CND should be used as the background if it dominates over the CMB; using the following procedure, we determined that this is not the case and that the CMB is an adequate background radiation field.  We compared the core continuum measurements of \citet{Krips:2006} to the CMB in our frequency range (88 to 363 GHz).  They measure the continuum emission as a power-law with frequency due to optically thin synchrotron radiation from the central jet source, which is embedded within the CND but smaller than the full 4\as\ diameter.  We use their measured emission sizes at 1 mm and 3 mm to estimate a power-law relating source size to frequency to convert from flux density to specific intensity.  However, the property which truly matters to the background radiation field is the mean intensity, which is not equivalent to the specific intensity because the size of the continuum emission is smaller than the full extent of the 4\as\ CND.  Therefore, the mean intensity is dependent upon the angular extent of the central source, which is a function of the frequency-dependent size of the source and the distance in the CND from it.  We use the same relationship between source size and frequency as mentioned above and determine an average angular size of emission throughout the CND, which we use to calculate the mean intensity.  We find that the mean intensity of the CMB radiation is comparable to that of our core + CMB model, and the difference in the predicted line intensities using either background model (CMB alone or core + CMB) is apparent in the RADEX results only at the lowest temperatures and densities; the overall likelihood results using grids with or without this background intensity added to the CMB are essentially indistinguishable.  However, we note that future studies using higher-J transition lines may still need to consider this background, which dominates over the CMB above 500 GHz.  In summary, the CMB + continuum emission is for the most part indistinguishable from the CMB alone, and so we use just the CMB for our background radiation field.

The input parameters required for RADEX are blackbody temperature of the background radiation (for our purposes, $T_{CMB} = 2.73$ K), kinetic temperature ($T_{kin}$), molecular hydrogen number density (assumed to be the only collision partner, $n(H_2)$), column density of the molecule ($N_{mol}$), and the width of the molecular lines.  Our calculations are all computed per unit linewidth because this is the physically relevant quantity that determines the optical depth scale; we later scale this value by the linewidth in our likelihood analysis.  Furthermore, RADEX assumes that the emission region fills the entire beam, but our likelihood analysis also introduces a beam filling factor $\Phi_A$ to account for clumpiness in the gas.

\begin{deluxetable}{c c c}
   \tablewidth{0pt}
   \tablecaption{RADEX Model Parameters and Ranges\label{table:radex}}
   \tablehead{
   \colhead{Parameter} & \colhead{Range} & \colhead{\# of Points}
	}
   \startdata
$T_{kin}$ [K] & $10^{0.7} - 10^{2.7}$ & 45\\
$n(H_2) $ [cm$^{-3}$] & $10^{2.5} - 10^{8.5}$ &  45\\
$N_{CS}$ [cm$^{-2}$] & $10^{10} - 10^{19}$ &  40\\
$X_{HCO^+}/X_{CS}$ & $10^{-1.5} - 10^{1.5}$ &   17\\
$X_{HCN}/X_{CS}$ & $10^{-1.0} - 10^{2.0}$ &  17\\
$X_{HNC}/X_{CS}$ & $10^{-1.5} - 10^{1.5}$ &  17\\
$\Delta V$ [km s$^{-1}$] & 1.0 &  fixed\\
$T_{background}$ [K] & 2.73 &  fixed
   \enddata
\tablecomments{All parameters are sampled evenly in log space.  $N_{CS}$ is given because it is defined as the primary species, see \S \ref{sec:RADEX}.}
\end{deluxetable}
We used RADEX to create a grid of line intensities for a range of $T_{kin}$, $n(H_2)$, and $N_{mol}$.  RADEX only treats one molecular species at a time, therefore in order to conduct a multi-species analysis, we first created a RADEX grid for one species, the primary species (CS).  We next created grids for all other (secondary) species, in which we also explore their relative abundances to the primary species.  In calls to RADEX, the column density was the product of the primary species' column density and the secondary species' relative abundance.  Our analysis used a relatively coarsely sampled but wide ranging grid, detailed in Table \ref{table:radex}.  Though each RADEX grid is created separately, the presence of multiple species at once is simultaneously considered in the likelihood analysis.

\subsection{Bayesian Likelihood Analysis}\label{sec:bayes}

We next compare the calculated line intensities to the measurements, detailed in Table \ref{table:likeflux}.  We add calibration error to the measurements in quadrature with the line uncertainties, assuming independent Gaussian uncertainties; this assumption is discussed in \S \ref{sec:sys}.

Given a set of measurements $\bm{x}$ and model parameters $\bm{p}  = (N_{CS},n(H_2),T_{kin},\Phi_A, \bm{X}_{mol} / \bm{X}_{CS})$, the Bayesian likelihood of the model parameters given the measurements is

\begin{equation}P(\bm{p} | \bm{x}) = \frac{P(\bm{p}) P(\bm{x} | \bm{p})}{P(\bm{x})},\end{equation}

\noindent where $P(\bm{p})$ is the prior probability of the model parameters (see \S \ref{sec:prior}), $P(\bm{x})$  normalizes, and $P(\bm{x} | \bm{p})$ is the probability of obtaining the observed data set given that the source follows the model described by $\bm{p}$, which is the product of Gaussian distributions in each observation,

\begin{equation}P(\bm{x} | \bm{p}) = \prod_i \frac{1}{\sqrt{2 \pi \sigma_i^2}} \rm exp \it \bigg[- \frac{(x_i - I_i (\bm{p}))^2}{2 \sigma_i^2} \bigg] \end{equation} 

\noindent Above, $\sigma_i$ is the standard deviation of the observational measurement for transition $i$ and $I_i({\bf p})$ is the RADEX-predicted line intensity for that transition and model.  The product is carried out over all of the molecular species considered.

\subsection{Prior Probabilities}\label{sec:prior}

The prior probability allows us to include previously known information about NGC 1068 in our likelihood analysis.  We created a ``binary" prior in which all physical situations were assigned $P(\bm{p}) = 1$, and all geometrically unphysical situations were assigned $P(\bm{p}) = 0$.  There were 4 criteria that each point in our model grid needed to satisfy in order to be deemed physically plausable.

\paragraph{Tau} At high optical depths, RADEX is unreliable because the cloud excitation temperature can become too dependent on optical depth with large column densities.  Therefore, we only include results with $\tau \leq 100$ as recommended by the RADEX documentation.  Most of the grid points excluded by this prior are at very high column density; in fact, the likelihood results are nearly identical with or without this prior condition.  The average optical depths for the columns that we find for our likelihood solutions are generally less than 10. 

\paragraph{Total mass} The mass in the emission region ($M_{region}$) should not exceed the dynamical mass of the galaxy, so we require that

\begin{equation}\label{eqn:mass}\frac{A_{region} N_{mol} \Phi_A 1.5 m_{H_2}}{X_{mol}} \leq M_{dyn}.\end{equation}

We are assuming the emission region is 312 pc wide (see \S \ref{sec:intro}), so $A_{region}$ is the corresponding area.  $N_{mol} \Phi_A$ is the beam-averaged column density, $m_{H_2}$ is the mass of the hydrogen molecule, the factor of 1.5 accounts for helium and other heavy elements, and $X_{mol}$ is the abundance of the molecule relative to hydrogen, $n_{mol}/n_{H_2}$.  

\citet{Usero:2004} conducted previous LVG simulations to determine the relative chemical abundances of many of the molecules discussed here; they found $X_{CS}$ to be $2.0 \times 10^{-8}$ in the East knot and $1.6 \times 10^{-8}$ in the West knot.  These values were derived assuming $X_{CO} = 8 \times 10^{-5}$.  Our beam does not distinguish between the two regions, so as a conservative limit we use the higher value.  For any grid point of a given column density and filling factor, using a higher value of $X_{mol}$ results in a lower estimate of the mass, meaning that fewer grid points overall will exceed the dynamical mass cutoff.  

The dynamical mass can be estimated as 

\begin{equation}M_{dyn} \approx \frac{2 \sigma_*^2 R}{G},\end{equation}

\noindent where $\sigma_*^2$ is the projected (1D) stellar velocity dispersion over an aperture of radius R, and the 2 accounts for 2D motion.  \citet{Oliva:1995} measured a velocity dispersion of 161 $\pm$ 20 \kms\ in a 4.4\as\ beam, which closely approximates the size of the region which we are studying.  This yields a dynamical mass of 2 $\times 10^9$ \ms, which we use as an upper limit for the mass constraint.

\paragraph{Velocity Gradient} We can estimate the observed velocity gradient of individual molecular clouds, and require that it be at least the minimum gradient required for virialization.  Using the expressions for these two quantities,

\begin{equation}\label{eqn:dvdr}
\bigg(\frac{\rm d v}{\rm d r}\bigg)_{obs} = \frac{\Delta v \ n_{H_2} X_{CS}}{N_{CS}}
\end{equation}

\noindent and

\begin{equation}
\bigg(\frac{\rm d v}{\rm d r}\bigg)_{vir} = \bigg(\frac{4}{3} \pi \alpha G \mu m_{H_2} n_{H_2}\bigg)^{1/2}, 
\end{equation}

\noindent where $\alpha = 1$, $\mu = 1.5$, and $m_{H_2}$ is the mass of the hydrogen molecule, we can place the requirement that

\begin{equation}
K_{vir} = \frac{(\rm d v / d r)_{obs}}{(\rm d v / d r)_{vir}} \geq 1.
\end{equation}

\paragraph{Column length} Finally, the length of the column ($L_{col}$) of the primary species should not exceed the length of the entire molecular region.  We assume that the length in the plane of the sky is the same as that orthogonal to the plane of the sky.  For the 4\as\ region we round the length of the region to 350 pc for the purposes of creating an upper limit.

\begin{equation}\frac{N_{mol}}{n(H_2) \sqrt{\Phi_A} X_{mol}} \leq L \end{equation}

\noindent We find that the length prior has the greatest effect on the likelihood results by excluding the lowest molecular hydrogen density and highest column density solutions.  Though it is the combination of these two parameters that matters in the length prior, the final marginalized likelihood results generally exclude solutions below $\sim 10^{3.5}$ cm$^{-3}$ in density and above $\sim 10^{16.5}$ cm$^{-2}$ in CS column density that would otherwise be allowed without this prior.

In summary, the prior probability is defined as:

\begin{equation}
P(\bm{p}) = \begin{cases} 1&\mbox{; } \tau \leq 100 \mbox{, } M_{region} \leq M_{dyn} \mbox{, }\\ &\ \ K_{vir} \geq 1 \mbox{, } L_{col} \leq L \\
            0&\mbox{; otherwise}\end{cases}
\end{equation}

\noindent The parameters that are assumed for the likelihood analysis are summarized in Table \ref{table:likeparams}.

\begin{deluxetable*}{c r c c c c c c c}
   \tabletypesize{\footnotesize}
   \tablecaption{Observed Line Intensities Used in Likelihood Analysis\label{table:likeflux}}
   \tablehead{
    &  &  &  &  &  & \colhead{Beam} & \colhead{Beam Corrected} & \\
   \colhead{Species} & \colhead{Transition} & \colhead{$\nu_{rest}$} & \colhead{$E_u/k_B$} & \colhead{$n_{crit}$\tablenotemark{a}} & \colhead{Intensity} & \colhead{FWHM} & \colhead{Intensity\tablenotemark{b}} & \colhead{Ref}\\
   \colhead{} &\colhead{}  & \colhead{[GHz]} & \colhead{[K]} & \colhead{[cm$^{-3}$]} & \colhead{[\kkms]} & \colhead{[\as]} & \colhead{[\kkms]} & \colhead{}
   }
\startdata
CS & J = 3 $\rightarrow$ 2 & 146.97 & 14.11 & 1.3 $\times 10^6$ & 9.5 $\pm$ 1.4 & 16.0 & 162 $\pm$ 34 & 1\\
   & 4 $\rightarrow$ 3 & 195.95 & 23.51 & 3.1 $\times 10^6$ & 1.56 $\pm$ 0.43 & 37.1 & 136 $\pm$ 40 & 2\\
   & 5 $\rightarrow$ 4 & 244.94 & 35.27 & 6.9 $\times 10^6$ & 1.23 $\pm$ 0.23 & 30.8 &  74 $\pm$ 16 & 2\\
   & 6 $\rightarrow$ 5 & 293.91 & 49.37 & 1.2 $\times 10^7$ & 1.17 $\pm$ 0.39 & 26.3 &  52 $\pm$ 18 & 2\\
   & 7 $\rightarrow$ 6 & 342.88 & 65.83 & 2.0 $\times 10^7$ & 1.4 $\pm$ 0.5 & 14.0 &  18.6 $\pm$ 7.2 & 3\\

\hline
HCO$^+$ & J = 1 $\rightarrow$ 0 &  89.19 &  4.28 & 2.2 $\times 10^5$ & 14.6\tablenotemark{c} $\pm$ 0.2 & 29.5 & 809 $\pm$ 122 & 4\\
        & 3 $\rightarrow$ 2 & 267.56 & 25.68 & 3.8 $\times 10^6$ & 5.58 $\pm$ 0.35 & 29 & 299 $\pm$ 35 & 2 \\
        & 4 $\rightarrow$ 3 & 356.73 & 42.80 & 9.1 $\times 10^6$ & 3.8 $\pm$ 0.5 & 14 & 50 $\pm$ 10 & 5\\

\hline
HCN & J = 1 $\rightarrow$ 0 &  88.63 &  4.25 & 3.2 $\times 10^6$ & $10 \pm 1.7$ & 44 & 1220 $\pm$ 177 & 6\\
    & 3 $\rightarrow$ 2 & 265.89 & 25.52 & 5.2 $\times 10^7$ & $8.5 \pm 0.26$ & 29 & 455 $\pm$ 48 & 2\\
    & 4 $\rightarrow$ 3 & 354.51 & 42.53 & 1.2 $\times 10^8$ & $13.9 \pm 1.6$ & 14 & 184 $\pm$ 35 & 5\\

\hline
HNC & J = 1 $\rightarrow$ 0 &  90.66 &  4.35 & 3.6 $\times 10^6$ & $11.4 \pm 0.7$ & 25 & 457 $\pm$ 74 & 7\\
    & 3 $\rightarrow$ 2 & 271.98 & 26.11 & 5.8 $\times 10^7$ & $2.15 \pm 0.21$ & 28 & 108 $\pm$ 15 & 2\\
    & 4 $\rightarrow$ 3 & 362.63 & 43.51 & 1.4 $\times 10^8$ & $2.7 \pm 0.3$ & 14 & 35.8 $\pm$ 6.7 & 5
\enddata
	\tablenotetext{a}{Calculated at 60 K for HCN and HNC, 70 K for CS and \hco from the Leiden Atomic and Molecular Database \citep{Schoier:2005}, where $n_{crit} = A_{ul}/\gamma_{ul}$.}
	\tablenotetext{b}{See \S \ref{sec:results} for details.  Also includes 10\% calibration errors for Z-Spec and 15\% for others.}
	\tablenotetext{c}{Later multiplied by $0.7 \pm 0.1$ for the Preferred model, see \S \ref{sec:fsb} for details.  Other lines are corrected for other models by factors described in the aforementioned section of the text.}
	\tablerefs{
(1) \citet{Mauersberger:1989b}, (2) Z-Spec (this work), (3) \citet{Bayet:2009}, 
(4) \citet{Krips:2008}, (5) \citet{Perez:2009}, (6) \citet{Perez:2007}, (7) \citet{Huttemeister:1995}.
	}
\end{deluxetable*}

\begin{deluxetable}{c c c}
\tablewidth{0pt}
\tablecaption{Likelihood Parameters Used\label{table:likeparams}}
\tablehead{\colhead{Parameter} & \colhead{Value} & \colhead{Units}}
\startdata
Line width\tablenotemark{a} &  240 & [km s$^{-1}$]\\
Abundance ($X_{CS}/X_{H_2}$)\tablenotemark{b} & 2.0 $\times 10^{-8}$ & \\
Angular Size Scale\tablenotemark{c} &  78 & [pc/\as]\\
Source size\tablenotemark{c} &  4.0 & [\as]\\
Length Limit\tablenotemark{b} & 350 & [pc]\\
Dynamical Mass Limit\tablenotemark{b} & 2 $\times 10^{9}$ & [\ms]
\enddata
	\tablenotetext{a}{Used for scaling the line intensities, see \S \ref{sec:fitting}.}
	\tablenotetext{b}{Used for prior probabilities, see \S \ref{sec:prior}.  Relative abundance is for the primary species, CS, relative to H$_2$.}
	\tablenotetext{c}{Used for prior probabilities (source size also used for scaling to correct for beam dilution), see \S \ref{sec:intro}.}
\end{deluxetable}

\subsection{Contamination of CND intensity by Starburst Ring}\label{sec:fsb}

\citet{Perez:2009} noted the $\sim$ 30\as\ diameter starburst ring could be contributing high-density tracer flux measureable by a large enough beam; because we are only interested in modeling the CND of NGC 1068, this extra flux would need to be removed from our calculations.  Though other single-dish measurements have sufficiently small beam sizes to avoid this problem at higher-J transitions, Z-Spec's beam is large enough (26-37\as\ FWHM) that we do need to take this into consideration.  They constructed a first-order estimate (their equations 1-4) of what they call the starburst contribution factor ($f_{SB}$) by comparing two measurements of the same line with different sized telescope beams.  The flux emergent from only the CND would be the total flux measured times (1-$f_{SB}$).  This estimate, however, yields high starburst contribution factors (around 50\%, with large uncertainties), which disagrees with currently available interferometric observations (such as those referenced in \S \ref{sec:intro}).

We attempted to calculate $f_{SB}$ for the Z-Spec measured lines by comparing them to other measurements in the scientific literature.  In this paragraph, in parentheses after each citation, we quote the reported velocity-integrated line intensities in \kkms\ and conversion to Jy \kms\ based on the references' reported beam sizes (the total integrated flux density in Jy \kms\ is the appropriate quantity to compare between telescopes)\footnote{Note that this conversion is different than the beam-dilution correction described in \S \ref{sec:results}, which assumes a source size (and produces the Beam Corrected Intensity column of Table \ref{table:likeflux}).  In this section we are simply converting to flux density units.}.  The Z-Spec measurement to which we are comparing is reported in Table \ref{table:fit}.  In our calculations, we add 20\% calibration error into the line measurements (this is slightly more than used elsewhere in the paper because of the additional uncertainty of the assumptions that went into the estimate itself, such as Gaussianity of the beam).  For the HNC \jthree\ line, we compare to \citet{Perez:2007} (3.5 \kkms, 69 Jy \kms) and find that for our beam, $f_{SB} = 0.54 \pm 0.37$ ($0.30 \pm 0.31$ for their JCMT beam).  In \hco\ \jthree\ we can compare to \citet{Krips:2008} (7.6 \kkms, 40 Jy \kms), who measured this transition using a 9.5\as\ beam.  At such a small beam size, the first order estimate essentially forces our measurement to match theirs, with $f_{SB} = 0.85 \pm 0.05$.  HCN \jthree\ was measured by \citet{Krips:2008} (19.0 \kkms, 99 Jy \kms), \citet{Perez:2007} (22.7 \kkms, 425 Jy \kms), and \citet{Bussmann:2008} (6.01 \kkms, 313 Jy \kms) but we find inconsistencies when comparing to these three other observations.  The last two fluxes in Jy \kms\ are comparable to ours despite different beam sizes (implying that we are measuring the same emission, which must be concentrated in the center if all beams are measuring it); this yields a negative, non-sensical estimate of the flux from the starburst region.  We can compare to the smaller (9.5\as) beam of \citet{Krips:2008} again yielding our $f_{SB} = 0.76 \pm 0.07$.  CS yields 3 measured transitions in our band, only 1 of which (\jfive) we are able to compare to other measurements from \citet{Martin:2009} (3.3 \kkms, 16.2 Jy \kms) with a beamsize of 10\as.  Our $f_{SB}$ for this line is $0.72 \pm 0.1$.  

These comparisons yielded suspiciously high contribution factors from the starburst ring.  The fact that our measured $f_{SB}$ values for the \jthree\ transitions with medium-sized beams (29\as) are higher than those derived by \citet{Perez:2009} for \jone\ and much larger beams (44-55\as) adds to this suspicion.  Additionally, \citet{Usero:2004} also made estimates of the contamination from the starburst ring by comparing central pointing values to those offset from it and found little contamination for SiO \jthree, no more than 25\% contribution for SiO \jtwo, and no more than 30\% for H$^{13}$CO$^+$ and \hco\ \jone.  Furthermore, in the case of HCN \jthree, we are already consistent with two other measurements cited above without any correction factor.  A likely explanation is that the values to which are we comparing are too low, potentially due to errors in calibration or baseline continuum measurements.  Another possibility is that the azimuthal structure of the starburst ring (or the beams) is significantly affecting the calculation of $f_{SB}$, which assumes such symmetry.

If we were to correct our measurements using the starburst correction factor described above, we find that the resulting spectral energy distributions of the CND for these high density tracers cannot reasonably be described by a single-temperature component of gas given the reported \jfour\ intensities of \hco, HCN, and HNC \citep{Perez:2009}, which rise above our corrected \jthree\ intensities when corrected for source size (as described in \S \ref{sec:results}).  There are two possibilities:  1) the \jfour\ lines are primarily excited in a separate component of gas within the CND that is warmer than the gas primarily traced by the lower J lines, or 2) the \jthree\ lines to which we are comparing in order to derive $f_{SB}$ are too low, perhaps due to pointing or calibration errors.

Because we currently have an incomplete picture of the extent of the contribution to these line strengths from the starburst ring, we present results for 3 scenarios.  The first is our preferred model, whose results will be presented in full.  We only make one correction for starburst emission in this model.  The majority of the lines used in our analysis have critical densities higher than SiO \jthree\ ($n_{crit} = 9.6 \times 10^5$ cm$^{-3}$ at 60 K), with the only exception being \hco\ \jone.  Based on Usero's measurements (which found that SiO \jthree\ had no contribution from the starburst ring), we assume in our preferred model that all of our molecular transitions are only excited in the CND, except for \hco\ \jone\ which we decrease by 30\%.  This implies that the gas in the CND is denser than in the starburst ring.  We believe this is the best model at the current time, though future interferometric studies may present new information that modifies this scenario.

The next two scenarios are presented in the Figures \ref{fig:like1}, \ref{fig:like2}, \ref{fig:columns}, \ref{fig:seds}, and \ref{fig:tau}, but not in any tables, though we discuss the differences in \S \ref{sec:physconds}.  The scenario labeled ``1-0 $f_{SB}$" uses the $f_{SB}$ values of \citet{Perez:2009} for the HCN, \hco, and HNC \jone\ values ($f_{SB}$ = 0.56, 0.45, and 0.45, respectively).  The final scenario labeled ``$f_{SB}$, Cool" uses the same $f_{SB}$ values as in ``1-0 $f_{SB}$," but also uses the $f_{SB}$ for Z-Spec measurements as described earlier in this section (for HCN, \hco, and HNC \jthree\ and CS \jfive, $f_{SB}$ = 0.76, 0.85, 0.54, and 0.72, respectively)  Because this reduces the \jthree\ lines, we exclude the \jfour\ lines, assuming they may be tracing a separate, warmer component.  We also include one extra line in this model, HCN \jtwo\ from \citet{Krips:2008}, which had been excluded from the other scenarios because of its low value.  The spectral energy distributions in \S \ref{sec:disc} should clarify the cases presented.  To summarize, the preferred model reduces only the \hco\ \jone\ line by 30\% due to starburst contribution, ``1-0 $f_{SB}$" corresponds to reducing emission of the ground transitions of HCN, HNC, and \hco\ by about half (assuming the other half is actually emission from the starburst ring), and ``$f_{SB}$, Cool" additionally corrects for a large starburst contribution in the middle-J states and excludes higher-J states which may be tracing a warmer component.


\section{Modeling Results and Discussion}\label{sec:disc}

\begin{figure}[t]
\begin{center}
\includegraphics[width=\columnwidth]{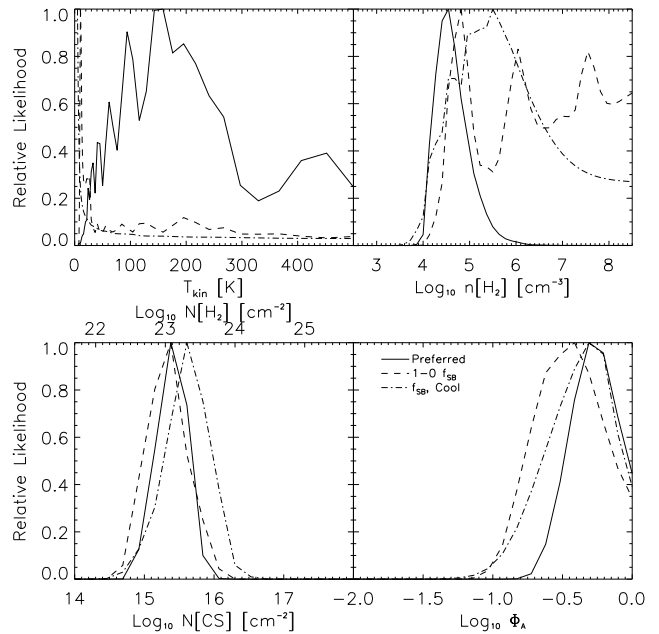}
\caption[Primary Parameter Likelihood Distributions]{Multiple-species probability distributions for the four primary parameters, $T_{kin}$, $n_{H_2}$, $N_{CS}$, and $\Phi_A$.  All are normalized such that the peak maximum likelihood value is equal to 1.  The hydrogen column density axis uses our assumed CS abundance of $2 \times 10^{-8}$ relative to H$_2$.  The different models as described in \S \ref{sec:fsb} are displayed with different linestyles: solid (``Preferred"), dashed (``1-0 $f_{SB}$"), and dash-dotted (``$f_{SB}$, Cool").}
\label{fig:like1}
\end{center}
\end{figure}

\begin{figure*}[t]
\begin{center}
\includegraphics[width=\textwidth]{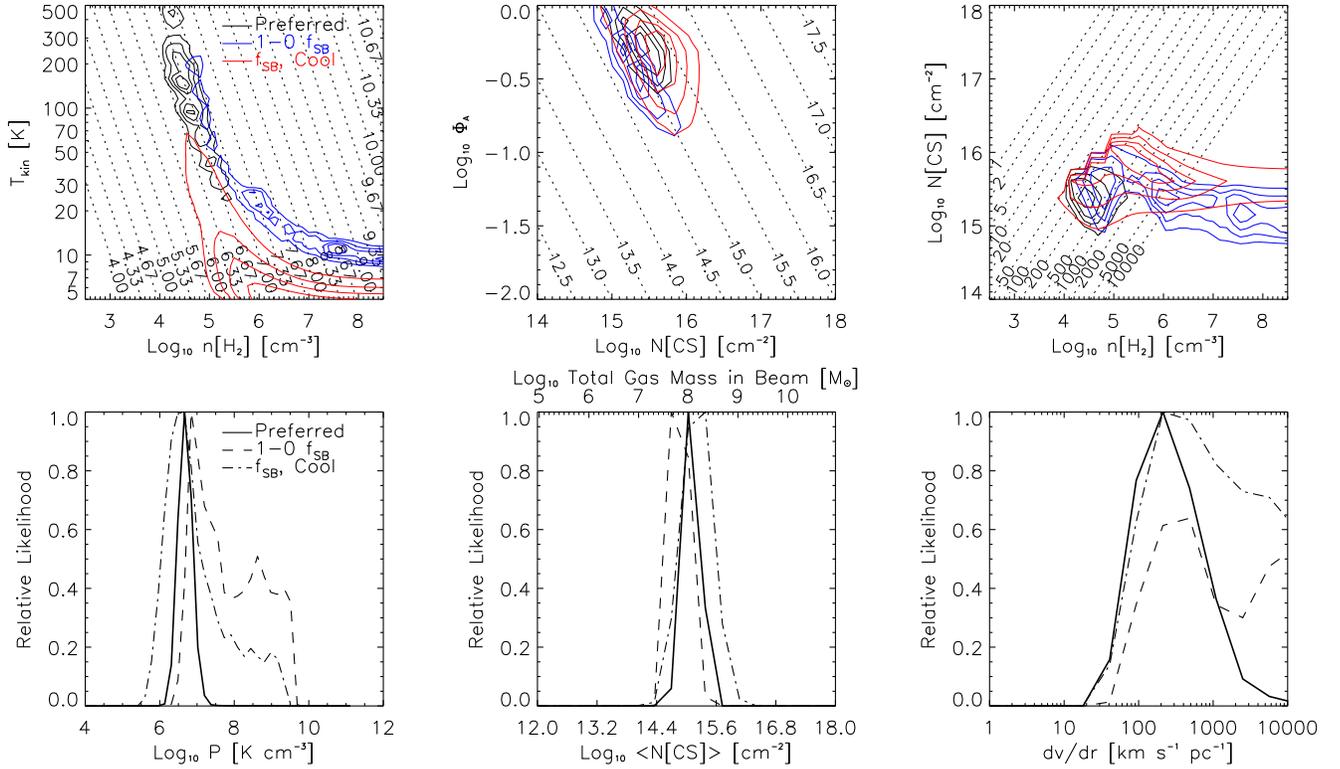}
\caption[Secondary Parameter Likelihood Distributions]{Multiple-species likelihood results: 2D distributions and supplementary parameters.  The top row contains 2D probability distribution contours for the parameters from Figure \ref{fig:like1}; each contour level represents 0.8, 0.6, 0.4, and 0.2 of the maximum of that distribution.  Diagonal lines on the contour plots correspond to values of the bottom parameter ($P$, $N(CS)$, and $dv/dr$ from left to right), which are derived from the primary parameters.  The different models as described in \S \ref{sec:fsb} are displayed with different colors (top) or linestyles (bottom): black/solid (``Preferred"), blue/dashed (``1-0 $f_{SB}$"), and red/dash-dotted (``$f_{SB}$, Cool").}
\label{fig:like2}
\end{center}
\end{figure*}

The raw result of our analysis is a matrix of likelihoods each characterized by a certain value of $T_{kin}$, $n(H_2)$, $N_{CS}$, $\Phi_A$, $X_{HCO^+}/X_{CS}$, $X_{HCN}/X_{CS}$, and $X_{HNC}/X_{CS}$.  To obtain a distribution for an individual parameter, we integrate the likelihood matrix over all other dimensions.  The integrated distributions for the first four parameters are shown in Figure  \ref{fig:like1}.  We can use these parameters to create distributions for 3 other supplementary parameters (bottom of Figure \ref{fig:like2}): thermal gas pressure, $P = n_{H_2} \times T_{kin}$, beam-averaged column density, $<N_{CS}> = \Phi_A \times N_{CS}$, and velocity gradient of individual clouds, Equation \ref{eqn:dvdr}.  From the beam-averaged column density we calculate the total molecular mass in the beam with Equation \ref{eqn:mass}, which is shown in the upper x-axis in the bottom middle panel of Figure \ref{fig:like2}.

\begin{figure}[t]
\begin{center}
\includegraphics[width=\columnwidth]{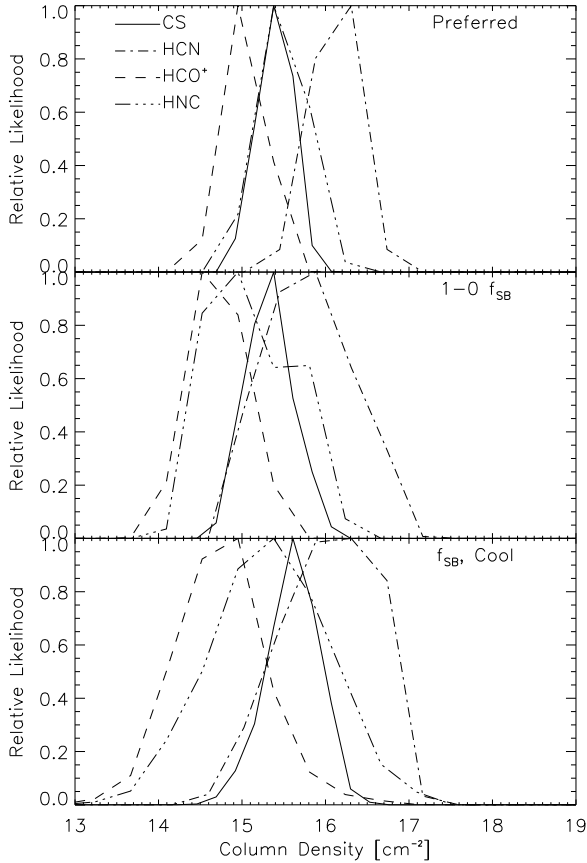} 
\caption[Multi-Species Column Densities]{Multiple-species column density likelihoods.  Each panel corresponds to one model as described in \S \ref{sec:fsb}.}
\label{fig:columns}
\end{center}
\end{figure}

\begin{figure}[t]
\begin{center}
\includegraphics[width=\columnwidth]{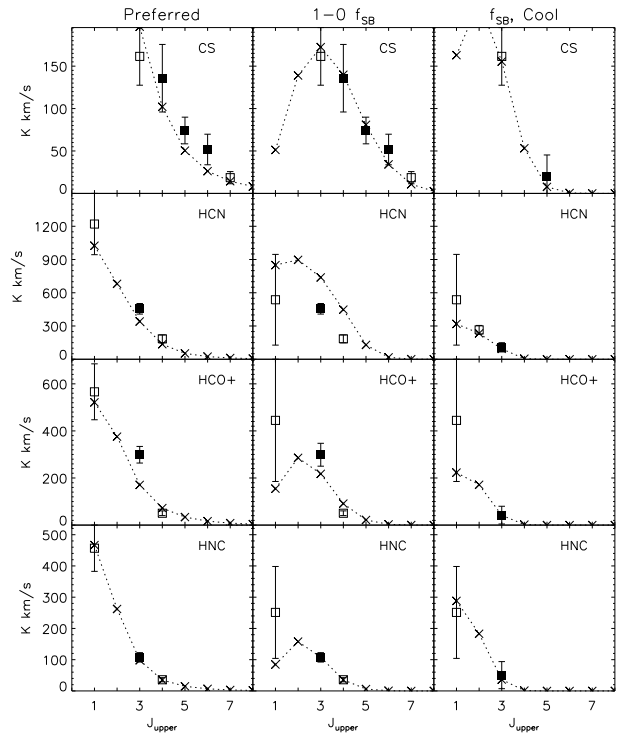} 
\caption[Spectral Energy Distributions]{Line spectral energy distributions.  Squares represent data points (see Table \ref{table:likeflux} for references used, filled squares indicate Z-Spec data points), X's represent the best fit model results.  All measurements are scaled for beam dilution, RADEX results are scaled by linewidth.  15\%\ calibration is added in quadrature for all measurements, except for Z-Spec, for which 10\%\ is added.  Each column is one model as described in \S \ref{sec:fsb}.}
\label{fig:seds}
\end{center}
\end{figure}

\begin{figure}[t]
\begin{center}
\includegraphics[width=\columnwidth]{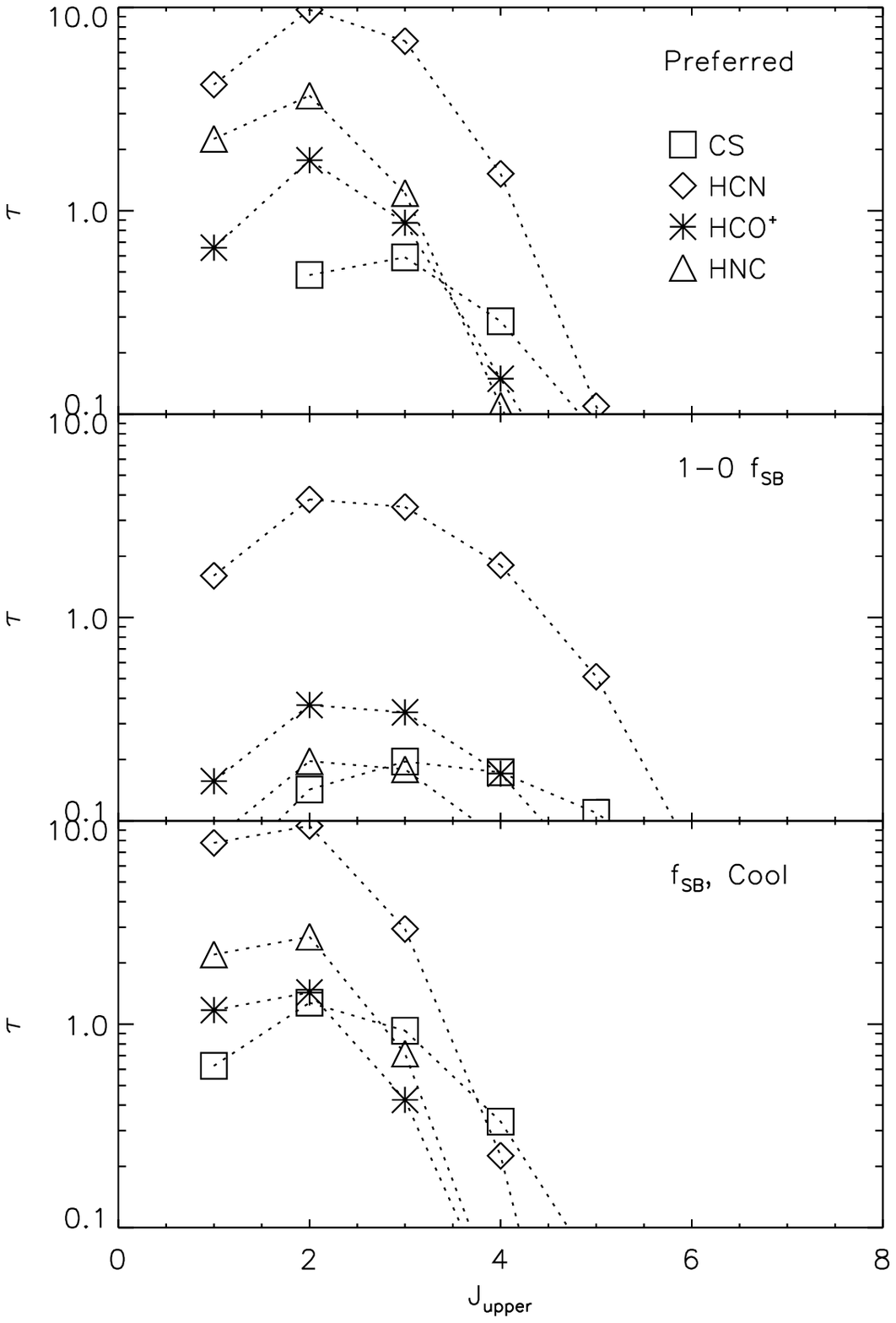} 
\caption[Optical Depths]{Optical depths for best-fit results.  The symbols correspond to molecules as follows: squares are CS, diamonds are HCN, asterisks are \hco, triangles are HNC.  Each panel corresponds to one model as described in \S \ref{sec:fsb}.  Optical depths are for the best-fit model results, as indicated in Figure \ref{fig:seds}.}
\label{fig:tau}
\end{center}
\end{figure}

The column densities for secondary molecules are shown in Figure \ref{fig:columns}.  The results for the distributions, marginalized over all other parameters, are in Table \ref{table:likeresults} (for the ``Preferred" model only).  The resulting spectral energy distributions and their optical depths are shown in Figures \ref{fig:seds} and \ref{fig:tau}.  Uncertainties cited in text are 1$\sigma$ unless otherwise stated.

\begin{deluxetable*}{c  c c c c}
\tablewidth{0pt}
\tablecaption{Multi-Species Radiative Transfer Results of Preferred Model\label{table:likeresults}}
\tablehead{
\colhead{Parameter} & \colhead{Median} & \colhead{1$\sigma$ Range} & \colhead{Max} & \colhead{4D Max} \\
}
\startdata
$T_{\rm kin}$ [K] & $117$ & $ 44$ - $231$ & $159$ & $159$\\
$n({\rm H_2})$ [cm$^{-3}$] & $ 3.4\times 10^{ 4}$ & $ 1.7\times 10^{ 4}$ - $ 8.7\times 10^{ 4}$ & $ 3.5\times 10^{ 4}$ & $ 2.6\times 10^{ 4}$\\
$N_{CS}$ [cm$^{-2}$] & $ 1.9\times 10^{15}$ & $ 1.0\times 10^{15}$ - $ 3.7\times 10^{15}$ & $ 2.4\times 10^{15}$ & $ 2.4\times 10^{15}$\\
$\Phi_{\rm A}$ & $ 0.48$ & $ 0.30$ - $ 0.72$ & $ 0.49$ & $ 0.62$\\
\hline
$P$ [K cm$^{-2}$] & $ 4.0\times 10^{ 6}$ & $ 2.3\times 10^{ 6}$ - $ 6.9\times 10^{ 6}$ & $ 4.7\times 10^{ 6}$ & $ 4.7\times 10^{ 6}$\\
$<N_{\rm CS}>$ [cm$^{-2}$] & $ 9.3\times 10^{14}$ & $ 4.2\times 10^{14}$ - $ 1.6\times 10^{15}$ & $ 1.1\times 10^{15}$ & $ 1.1\times 10^{15}$\\
d$v$/d$r$ [km s$^{-1}$ pc$^{-1}$] & $167$ & $51 - 581$ & $ 212$ & $ 212$\\
Mass in Beam [M$_\odot$] & $ 8.5\times 10^{ 7}$ & $ 3.8\times 10^{ 7}$ - $ 1.4\times 10^{ 8}$ & $ 1.0\times 10^{ 8}$ & $ 1.0\times 10^{ 8}$\\
\hline
$X_{\rm HCN}/X_{\rm CS}$ & $  4.8$ & $  3.1$ - $  7.6$ & $  4.9$ & $  4.9$\\
$N_{\rm HCN}$ [cm$^{-2}$] & $ 7.9\times 10^{15}$ & $ 2.8\times 10^{15}$ - $ 2.3\times 10^{16}$ & $2.0\times 10^{16}$ & $ 2.0\times 10^{16}$\\
$X_{\rm HCO^+}/X_{\rm CS}$ & $ 0.30$ & $ 0.23$ - $ 0.47$ & $ 0.42$ & $ 0.42$\\
$N_{\rm HCO^+}$ [cm$^{-2}$] & $ 7.2\times 10^{14}$ & $ 2.6\times 10^{14}$ - $ 1.4\times 10^{15}$ & $ 8.9\times 10^{14}$ & $ 8.9\times 10^{14}$\\
$X_{\rm HNC}/X_{\rm CS}$ & $ 1.07$ & $ 0.79$ - $ 1.66$ & $ 1.55$ & $ 1.55$\\
$N_{\rm HNC}$ [cm$^{-2}$] & $ 2.0\times 10^{15}$ & $ 6.9\times 10^{14}$ - $ 5.5\times 10^{15}$ & $ 2.4\times 10^{15}$ & $ 2.4\times 10^{15}$
\enddata
\tablecomments{1D max refers to the maximum value of the integrated parameter distribution.  4D max refers to the value of that parameter at the best fit solution.}
\end{deluxetable*}

\begin{deluxetable}{c c}
\tablecaption{Relative Abundances\label{table:abundances}}
\tablewidth{0pt}
\tablehead{\colhead{Abundance Ratio} & \colhead{Median Value}}
\startdata
$X_{\rm HCN}/X_{\rm HCO^+}$ & $15.8 \pm 10$ \\
$X_{\rm HNC}/X_{\rm HCN}$ & $0.22 \pm 0.15$ \\
$X_{\rm CS}/X_{\rm HCN}$ & $0.21 \pm 0.1$ \\
$X_{\rm HCN}/X_{\rm CO}$ & $  0.0012 \pm 0.0005$ \\
$X_{\rm CS}/X_{\rm CO}$ & $2.5 \times 10^{-4}$
\enddata
\tablecomments{All values are derived from Table \ref{table:likeresults}.  CS and CO use assumed abundances to H$_2$ of $2 \times 10^{-8}$ and $8 \times 10^{-5}$, respectively.  Uncertainties are representative.}
\end{deluxetable}

\subsection{Physical Conditions}\label{sec:physconds}

Warm gas ($44-231$ K) is implied in the nuclear disk.  Because the temperature distribution is so broad, it is possible that a single-temperature model is not adequate for the CND of NGC 1068.  It is important to note that the temperature and the density are degenerate; a high-temperature, low-density model may reproduce the data equally as well as a low-temperature, high-density model.  Therefore, it is not surprising that both our temperature and density distributions are somewhat broad.  However, for the preferred model, low temperatures ($< 50$ K) are disfavored, which corresponds to the upper limit on the density likelihood distribution.  The density is constrained to $10^{4.2} - 10^{4.9}$ and the product of the temperature and density (pressure, Figure \ref{fig:like2}) is better constrained with Log(P $[\unit{K \ cm^{-3}}]) = 6.60 \pm 0.24$).  We note that decreasing the \hco\ \jone\ flux can bias the results towards higher pressures (as can be seen in the higher pressures of the other models, bottom left panel of Figure \ref{fig:like2}); however, as will be discussed in \S \ref{sec:sys}, \hco\ does not dominate the results.

Previous models have assumed temperatures of around 50 K or 80 K, but we have shown that the gas (or at least a significant component of it) likely traces a higher temperature ($\sim 159$ K) region.  This distinction matters because assuming a lower temperature necessarily, and perhaps incorrectly, increases the density that is required to find solutions.  \citet{Krips:2008} found 2 best fit models for NGC 1068, one low-temperature/high-density and one high(unconstrained)-temperature/low density.  By examining the relative likelihoods over a large parameter space, we have demonstrated that there is broader support for their second solution, which matches well with our measurements.  At the least, it is clear that a high-temperature component is needed to adequately explain the measured spectral line energy distributions. 

The column densities and relative abundances of the molecules are well constrained, as can be seen in Figure \ref{fig:columns} and Table \ref{table:likeresults}.  Moreover, the column densities are fairly consistent among the different models, though CS and HNC ``switch" places relative to one another in bottom two models/panels in Figure \ref{fig:columns}; however, their distributions still overlap in all models.  Another strongly constrained parameter from these models is the total mass in the beam, bottom middle panel of Figure \ref{fig:like2}, at $10^8$ \ms.

We have referred all of our measurements to a 4\as\ source size and find, with that referred size, an area filling factor of $0.48^{+0.24}_{-0.18}$ (for the preferred model), which corresponds to a volume filling factor $0.33^{+0.28}_{-0.17}$ (assuming spherical symmetry and therefore $\Phi_V = \Phi_A^{3/2}$).  This implies some clumpiness in the gas.  Our inferred mass in this 4\as\ area is 1.0 $\times 10^8$ \ms, very similar to \citet{Israel:2009}'s estimate of 1.2 $\times 10^8$ \ms as inferred from CO.

The derived velocity gradient is much greater than what may be estimated by simply dividing the total observed velocity linewidth by the length of the CND ($\sim$ 1 \kmspc).  This indicates that the CND is likely composed of clumpy, turbulent structures.  Such high velocity gradients have been found in studies of the gas in the NGC 253 starburst nucleus \citep{Bradford:2003,Hailey-Dunsheath:2008} and can arise naturally due to the geometry and kinematics in compact circumnuclear environments (e.g. \citet{Bradford:2005}).  

In our preferred model, we have only corrected \hco\ \jone\ for contributions from the starforming ring (see \S \ref{sec:fsb} for justification, and \S \ref{sec:sys} for discussion of the effects of this choice).  If we use the $f_{SB}$ correction factors of \citet{Perez:2009} (``1-0 $f_{SB}$", dashed line in Figures \ref{fig:like1} and \ref{fig:like2}), as well as those for the \jthree\ lines that we can derive by comparing Z-Spec's measurements to other data (``$f_{SB}$, Cool", dashed-dotted line in Figures \ref{fig:like1} and \ref{fig:like2}), the likelihood results (especially for temperature and density) are dramatically different.  The temperature peaks at 5 K, but is unconstrained (constant relative likelihood) at higher temperatures.  As a result, the density distribution is significantly broader and unconstrained even above $10^7$ cm$^{-3}$.  The resulting SEDs (Figure \ref{fig:seds}), after likely overcorrecting for contribution from the starburst ring, do not seem to be describing physically reasonable parameters.  Even if we only use their derived \jone\ correction factors and leave all other measurements uncorrected, the temperature still peaks at a very low value (12 K) and is unconstrained along with the density.  However, as mentioned, the column densities are still fairly well constrained, and given this constraint the models show a trade-off between $\Phi_A$ (decreasing with each model) and density (increasing with each model).  We believe the SEDs that we are modeling, where only \hco\ \jone\ needs a correction for contribution from the ring, are the most reasonable considering interferometric maps described in \S \ref{sec:intro} and also produce more physically reasonable results than if we apply extra corrections.  However, we recognize that there may still be some level of contamination from the starburst ring that we cannot properly calculate, and therefore the results may represent a combination of the CND and ring, but should be dominated by the CND.  The effects of the assumed $f_{SB}$ values are further discussed in the next section.

\subsection{Systematic Effects}\label{sec:sys}

It is important to note a few systematic effects that exist within our likelihood analysis.  We have assumed that the measurements all have independent Gaussian uncertainties.  Such a treatment would only be completely valid in the case that the calibrations for each line measurement were independent; however, the Z-Spec line fluxes have correlated calibration uncertainties.  Additionally, multiple lines come from the IRAM 30-meter telescope and the James Clerk Maxwell Telescope (JCMT), and the lines from each individual telescope have correlated calibration uncertainties.  Because it would be difficult to assign relative calibration errors to the various line measurements (they are not reported uniformly), we do not attempt to treat the calibration errors as a separate systematic error.  Rather, we simply assess the impact of the inclusion of the assumed relative calibration errors in quadrature with the random errors.  Doubling the calibration errors results in likelihood distributions that widen (e.g. by a factor of 1.4 for pressure at the 1-sigma range), but the median values change insignificantly.  This indicates that line fluxes determine the molecular level populations and the sizes of the assumed relative calibration errors contribute to the likelihood widths.  (Of note, larger calibration errors reduce the jaggedness of the temperature likelihood distribution in Figure \ref{fig:like1}.  This implies that the grid point spacing is slightly sparse relative to the measurement uncertainties, but since the mean and overall shape do not change, there are no major consequences for our conclusions.)

We use the linewidth to scale the intensities by first dividing each velocity-integrated intensity by the assumed linewidth to compare directly to the RADEX predictions of Raleigh-Jeans temperature for a given column density per unit linewidth.  Then, in order to plot the likelihood for the total column density, we multiply column density per unit linewidth by the linewidth.  We assumed the same linewidth (240 \kms) for each molecular line transition based on the CO \jtwo\ line of \citet{Israel:2009}, though there are slight variations by molecular species.  For example, Tables 3 of \citet{Perez:2007} and 2 of \citet{Perez:2009} illustrate the linewidth for some of the molecular transitions used here; many are broken in multiple components, and some single-component measurements have linewidths as high as 275 \kms\ (HCN \jthree) and as low as 166 \kms\ (\hco\ \jfour).  Changing the assumed linewidth modifies the likelihood results for the column density.  Using a linewidth of 166 instead of 240 \kms\ reduces median CS column density by 25\%, and using a linewidth of 270 instead of 240 \kms\ increased the median CS column density by less than 1\% (the distributions for the area filling factor shift in the opposite manner from column density in similar proportions).  Additionally, the supplemental parameter of velocity gradient would be affected by using a different linewidth (in proportion with that linewidth, i.e. dv/dr would be 30\% smaller if the linewidth were 30\% smaller).  Thus, given this range of possible values, the systematic errors due to linewidth assumptions are less than 30\%.

The CS abundance was assumed; this does not affect the relative abundances of the secondary to the primary species, but would affect the conversion to molecular hydrogen.  Both the total mass in the beam and the velocity gradient depend (in inverse ways) on X$_{CS}$.  The fact that our estimate for the gas mass agrees with \citet{Israel:2009} suggests that the value we have adopted for X$_{CS}$ is about right.  Finally, X$_{CS}$ is used in the prior probabilities to determine if individual grid points violate the maximum dynamical mass, length, or velocity gradient criteria.  If X$_{CS}$ is too high/low then more/fewer grid points are allowed.  The main effect of the prior probability is to cut the lowest density and highest column density points from consideration in our analysis.

This analysis necessarily assumes that all the molecular lines trace the same gas.  Detailed line profiles indicate that not all lines necessarily trace the same kinematic structures (i.e. \citet{Perez:2007}); however, this analysis is an attempt to examine the bulk properties of the CND.  We investigated this claim by also examining the likelihood distributions created by modeling individual molecules one at a time.  These single-species models are less constrained than the multi-species model (as one would expect by using fewer molecular lines), but show some mutually consistent results.  The temperature and density distributions of HNC are most similar to the multi-species models.  HCN and CS show similar profiles, and though their best-fit solution requires temperatures of $\sim$ 160 and 80 K, respectively, the marginalized distribution also shows a pronounced peak at lower ($\sim$ 10 K) temperatures.  Because of these allowed lower-temperature solutions, the density distribution for these two molecules peaks slightly higher than the multi-species model and is less constrained at the high-density end, though their uncertainties are quite large; the 1$\sigma$ uncertainties for density of HCN and CS span 1.5 and 2.4 orders of magnitude, respectively.  \hco is the only molecule that, when modeled alone (including the 30\% correction to \jone), only demonstrates low-temperature/high-density solutions, though the density is also very unconstrained, as the 1$\sigma$ range spans 2.4 orders of magnitude.  When we model only HNC, HCN, and CS (excluding \hco), the results do not change significantly compared to when we include \hco, indicating that is not dominating the multi-species likelihood results, but may be tracing a separate component of gas.

Though we acknowledge that our treatment of the CND as one chemical/physical component is a simplifying assumption that is likely not strictly correct, our results still demonstrate fairly consistent results across multiple molecules.  We present these results with the understanding that within this large region, there are likely multiple chemical/physical components, but with single-dish measurements we are only capable of modeling and commenting on the bulk properties of the region.  These physical parameters describe the sum of the emission we see, but not the individual components.  ALMA will be key to resolving individual components within the CND.

Finally, we note that there are admitted uncertainties in the starburst contribution factors ($f_{SB}$) for our preferred model, which we have only corrected for \hco\ \jone, using $f_{SB} \sim 30$\%.  This correction serves to slightly increase the density while decreasing the temperature and filling factor, though the column density is not significantly changed.  The pressure distribution is only slightly increased (our model peaks at $4.7 \times 10^6$ K cm$^{-2}$, though it would peak at $3.1 \times 10^6$ K cm$^{-2}$ without the correction).  The mass distribution is slightly changed as well, as without the correction it would peak at $2 \times 10^8$ \ms, a factor of 2 higher than our model.  In total, the \hco\ \jone\ correction changes the distribution only a small amount relative to the uncertainties.  For the other models, we first note that the current values of $f_{SB}$ used are subject to systematic errors in their derivation, because the model is sensitive to both the beam profiles, which may not be purely axisymmetric Gaussians, and also to the relative calibration between the telescopes (which is notoriously difficult).  For any individual molecule, as we increase $f_{SB}$ (reducing the \jone\ intensity for HCN, HNC, and \hco), the best-fit solution's density increases and is less constrained, and the relative likelihood of higher temperature decreases.  In general, the filling factor also decreases.  Higher $f_{SB}$ scenarios require solutions which peak at the \jtwo\ line instead of \jone\ (see the middle column of Figure \ref{fig:seds}), which we cannot constrain given no \jtwo\ measurements.  Our slightly lower-density solutions may be a result of uncertainties in $f_{SB}$, though we have attempted to model the CND given all the currently available information that we have.  In the future, when $f_{SB}$ is better constrained through interferometric measurements, our models may require modification that will result in slightly higher-density solutions.  

\section{Diagnostics of the AGN Central Engine}\label{sec:diagnostics}

The results of our radiative transfer analysis provide diagnostic information about the energetics of the CND.  We first present a discussion of our derived relative molecular abundances and how they fit into the context of previous studies of the CND, especially its XDR nature.  We next present a separate analysis of line ratios specific to the Z-Spec band, which also supports the XDR scenario.  Finally, we speculate on the lifetime of the CND as determined by the black hole accretion rate and inferred molecular gas reservoir.

\subsection{Molecular Abundances and Possible Energy Sources for the Molecular Gas Excitation}\label{sec:abundances}

Some relative abundances of interest are presented in Table \ref{table:abundances}.  We particularly note the high relative abundance of HCN to \hco.  The median values from Table \ref{table:likeresults} yield [HCN]/[\hco] of $\approx$ 16.  This is higher than \citet{Usero:2004} (with ratios of 1.3 and 0.6 for the East and West knots, respectively) but more consistent with \citet{Krips:2008}, who required [HCN]/[\hco] of at least 10 in order to find a solution that matched measured values.  

There are three ways to produce such a high abundance.  HCN could be enhanced in a PDR by far-UV radiation from massive star formation.  However, there is no evidence for significant current star formation in the CND of NGC 1068.  The starburst activity detected by PAH emission likely contributes only 1\%\ of the total infrared luminosity \citep{Marco:2003} and the nuclear starburst is likely 200-300 Myr old \citep{Davies:2007}.

Oxygen depletion could also be responsible for decreasing the abundance of \hco, hence increasing the HCN/\hco\ ratio.  Oxygen underabundance would reduce the formation of CO, leaving more carbon for other carbonated species (e.g. HCN and HNC, \citet{Usero:2004,Sternberg:1994,Shalabiea:1996}).  Since \hco\ includes both carbon and oxygen, the effects of more available carbon but less available oxygen mean no net change in its abundance.  Therefore, this scenario would predict higher HCN/CO and HCN/\hco\ ratios.  This scenario was ruled out by \citet{Usero:2004} because it predicted a higher ratio of HCN/\hco\ than they found; however, our ratio (which agrees with \citet{Krips:2008}) matches better with the prediction from the cited \citet{Usero:2004} oxygen depletion model than their cited XDR model.  The same is true for the HCN/CO abundance ratio.  Because their oxygen depletion model is based on different physical conditions than we find here ($n_{H_2} = 1 \times 10^{5}$ cm$^{-3}$, $T = 20$ K, \citet{Ruffle:1998}) , we do not comment on the oxygen depletion scenario based on their models.

The final possibily is an XDR, which at lower densities (such as those inferred from our analysis, median of $10^{4.5}$) can explain the HCN/\hco\ ratio. In addition to the modeled molecular abundances, the measured line intensity ratios also lend evidence towards the XDR scenario, as shown in the following section.  We also found a ratio of column densities N(HNC)/N(HCN) $< 1$ (~ $0.3$), which can be found in XDR environments if the total column density is lower than $10^{24}$ cm$^{-2}$ \citep{Perez:2007}, which we infer through our likelihoods that it is.

Some other possibilites for the energetics of NGC 1068 have also been considered.  The first two have been ruled out; \citet{Krips:2008} pointed out that cosmic rays accelerated in supernova remnants are generally thought to increase \hco\ and decrease HCN abundances, which is the opposite of what we observe.  IR pumping affects both \hco\ and HCN, which would not explain the high HCN/\hco\ ratio.  However, \citet{Loenen:2008p3375} found that some XDR results can be explained also by a weak PDR with mechanical heating, and the high abundance of SiO may indicate the influence of shocks \citep{GarciaBurillo:2010}.

\subsection{Line Ratios: Strong Evidence of XDR Excitation of Molecular Gas}\label{sec:ratios}

In addition to radiative transfer and non-LTE excitation modeling, the line ratios of high density tracers can themselves be used as diagnostics for XDR environments.  Lines within the Z-Spec band benefit from our common calibration, making the ratios within our band ideal for these diagnostics.  Futhermore, even if the lines suffer from contamination from the starburst ring, if that contamination is of a similar amount for the lines we are comparing (i.e. they have similar values of $f_{SB}$), then their ratio is unaffected.  Some of the intensity ratios constructed using Z-Spec only are shown in Table \ref{table:ratios}.  Though our \hco/HCN \jthree\ ratio is higher than \citet{Krips:2008} ($0.38 \pm 0.07$) and our HCN/HNC slightly lower than \citet{Perez:2007} ($6.48 \pm 1.95$), the ratios cannot be concluded to be incompatible given the error bars.

\begin{deluxetable}{c c}
   \tablewidth{0pt}
   \tablecaption{Z-Spec Determined Line Ratios in NGC1068\label{table:ratios}}
   \tablehead{\colhead{Ratio} & \colhead{Value}}
   \startdata
HCN$_{3 \rightarrow 2}$/HNC$_{3 \rightarrow 2}$   & 4.2 $\pm$ 0.4 \\
HCO$^+_{3 \rightarrow 2}$/HNC$_{3 \rightarrow 2}$ & 2.8 $\pm$ 0.3\\
HCO$^+_{3 \rightarrow 2}$/HCN$_{3 \rightarrow 2}$ & 0.67 $\pm$ 0.05\\
CS$_{6 \rightarrow 5}$/CS$_{5 \rightarrow 4}$     & 0.70 $\pm$ 0.27\\
CS$_{5 \rightarrow 4}$/CS$_{4 \rightarrow 3}$     & 0.55 $\pm$ 0.18\\
CS$_{5 \rightarrow 4}$/HNC$_{3 \rightarrow 2}$    & 0.69 $\pm$ 0.15\\
CS$_{5 \rightarrow 4}$/HCO$^+_{3 \rightarrow 2}$  & 0.25 $\pm$ 0.05\\
CS$_{5 \rightarrow 4}$/HCN$_{3 \rightarrow 2}$    & 0.16 $\pm$ 0.03
   \enddata
\tablecomments{Errors do not include calibration error (estimated to be 10\%) because they are all measured within our Z-Spec band.}
\end{deluxetable}

\begin{figure}[t]
\begin{center}
\includegraphics[width=\columnwidth]{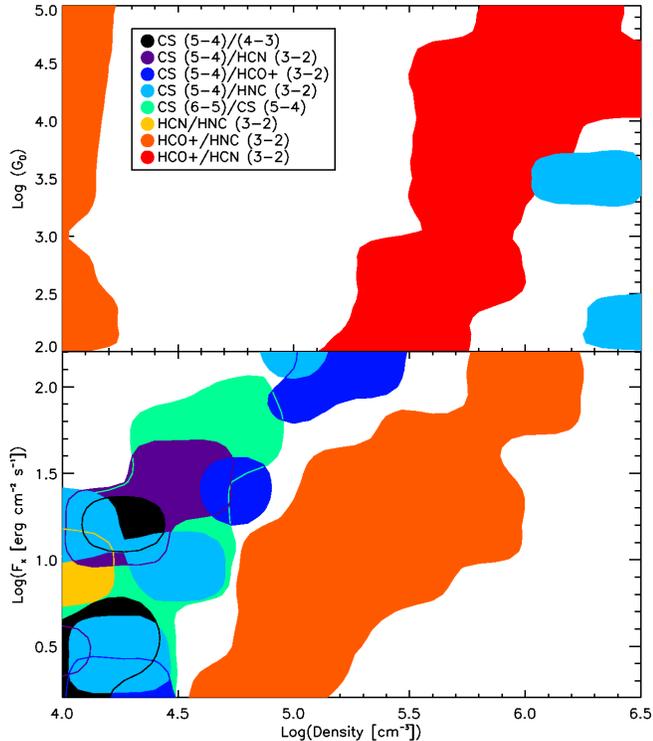}
\caption[2-Sigma Confidence Intervals for Z-Spec Line Ratios]{2$\sigma$ confidence intervals for Z-Spec line ratios.  These confidence intervals were constructed by comparing Z-Spec's measured line ratios (Table \ref{table:ratios}) with the PDR (top) and XDR (bottom) models of \citet{Meijerink:2005}.}
\label{fig:XDRPDR}
\end{center}
\end{figure}

\citet{Meijerink:2005} constructed grids of XDR and PDR models.  For our molecular lines of interest, the grids cover a range of densities ($10^4 - 10^{6.5}$ cm$^{-3}$), sizes (hence, column densities, $3 \times 10^{22} - 1 \times 10^{25}$ cm$^{-2}$), and radiation fields.  The PDR UV-radiation field is represented in multiples of the Habing flux ($G_0 = 1$ corresponds to $1.6 \times 10^{-3}$ erg cm$^{-2}$ s$^{-1}$) and the XDR X-ray field is simply given in erg cm$^{-2}$ s$^{-1}$.  Intensities for various J lines for these molecules are calculated at each grid point and contours of some line ratios are presented assuming cloud sizes of 1 pc (see Figures 14-17 in \citet{Meijerink:2007}).  Assuming this fixed cloud size means that the column density varies as a function of the density.  The line ratios within the Z-Spec band are not plotted explicitly in their publication, but the grids are available online such that we could make our own diagnostic plots\footnote{http://www.strw.leidenuniv.nl/$\sim$meijerink/grid/}.  We constructed 8 line ratios that are unique to Z-Spec's band, as presented in Table \ref{table:ratios}, and created contour plots for the XDR and PDR models for each ratio.  We then constructed a chi-squared statistic comparing our ratio to each grid point; the resulting 2-sigma confidence intervals (smoothed) for each of the 8 ratios are presented in Figure \ref{fig:XDRPDR}.

Of the 8 line ratios we present, 6 are consistent with the XDR model at the lower densities inferred by our radiative transfer modeling (between $10^4$ and $10^5$ cm$^{-3}$).  We note that the models do not go below $10^4$ cm$^{-1}$, so for those molecules which favor solutions at this density, the parameters should be considered upper limits.  The remaining 2 ratios are either consistent with XDR (\hco/HNC \jthree) or PDR (\hco/HCN \jthree) at higher densities.  The fact that the two ratios in disagreement with the others contain \hco\ may indicate something about the physical state of this particular molecule; for example, it may be tracing shock-excited material.  \citet{GarciaBurillo:2010} have found previous evidence of shocks in NGC 1068 using SiO, HNCO, and methanol (CH$_3$OH).  Our conclusions are similar to the aforementioned paper; we find some evidence of XDR influence, but also some possible influence of shocks.  The discrepancy of \hco\ may also lend support to the oxygen depletion scenario, because in order to match the XDR models at lower densities, both of these ratios would need to be significantly higher than we measure.  If HCN and HNC are being selectively enhanced over \hco, that could drive down these ratios, and hence explain why we see lower values than those predicted by the XDR models at densities of $\sim 10^{4.5}$ cm$^{-3}$.  If \hco\ is in fact also tracing gas undergoing a different excitation mechanism, we again consider the possibility that it should not be included in the radiative transfer analysis (as in \S \ref{sec:sys}).  When we exclude this molecule, the density peaks at $10^{5.0}$ instead of $10^{4.5}$ cm$^{-3}$.  Even with the slightly larger density, there is no additional support for the PDR model, given the arguments already presented that remain the same even when \hco\ is excluded from the analysis.  However, this is now a second, independent line of evidence (in addition to the discussion in \S \ref{sec:sys}) that suggests that \hco\ may be tracing a different gas component than other tracers. 

This analysis adds support to previous studies that have demonstrated the XDR nature of the CND of NGC 1068.  For example, \citet{Usero:2004} found support for the XDR model based on relative molecular abundances derived from LVG simulations (HCN/CO, CS/CO, HCO/\hco,CN/HCN).  \citet{Perez:2009} came to the same conclusion based on line ratios of \hco\ \jfour\ with lower J lines and HCN \jfour, and by the N(CN)/N(HCN) $\sim$ 1-4 column density ratio.  These studies are in addition to the \citet{GarciaBurillo:2010} conclusions in the previous paragraph.  Also, \citet{Maloney:1997p2554} demonstrated that the X-ray irradation in the CND is adequate for creating an XDR.

The agreement of the majority of our line intensity ratios with the XDR scenario at densities consistent with our radiative transfer modeling results provides strong support for an XDR and also demonstrates the utility of Z-Spec's broad bandpass and self-consistent line calibration across the band.

\subsection{Black Hole Accretion Rate and CND lifetime}

We used our derived total gas mass ($10^{8}$ \ms) to estimate the timescale for accretion onto the central black hole, based on the accretion rate inferred from the AGN luminosity.  We use the simple equation  

\begin{equation}
L = \eta \dot M c^2 , 
\end{equation}

\noindent where $\eta$ is the efficiency.  Using an accretion luminosity of $3.1 \times 10^{11}$ \ls \citep{Rigby:2009} (based on [OIV] as a calibrator) , the accretion rate is 0.2 \ms\ year$^{-1}$ assuming 10\%\ efficiency.  Estimating a timescale for accretion by $M_{CND}/\dot M$ yields $\sim$ 500 Myr.  This is only a few times the 200-300 Myr estimated age of the nuclear stellar core (the innermost 70 pc).  By this estimate, the most recent star formation activity has only taken place for a portion of the AGN's depletion timescale.  We note that the timescale of accretion is proportional to efficiency, which is not well known, but most likely we are underestimating the timescale rather than overestimating, considering that the efficiency is likely between 10-100\% \citep{Schartmann:2010} but not significantly lower than 10\%.  There are also uncertainties in the total gas mass, leaving it difficult to precisely compare the timescales.  

\citet{Davies:2007} found a correlation between the age of nuclear starburst and AGN luminosity; of their sample of 8 galaxies, NGC 1068 had one of the oldest starbursts and also is one of the most luminous AGN, accreting at a significant fraction of its Eddington luminosity.  It is interesting that the CND hosts a significant amount of dense gas that could be used for star formation, yet no significant star formation is currently happening.  \citet{Davies:2007} theorize that there is a delay between a starburst and AGN activity and that the starburst activity can help power the AGN.  Though the nuclear starburst began 200-300 Myr ago and has since effectively ended, its influence is still significant because it likely contributed to making NGC 1068 such a strong AGN, which now is the primary influence of the dense molecular gas as we saw in the XDR analysis of the previous section.

\section{Conclusions and Future Work}\label{sec:concl}

We have presented a new broadband spectrum of NGC 1068 obtained with Z-Spec, complementing the current literature with a set of common-calibration measurements of CO, its isotopes, and other high-density tracing molecules.  After extending the modeling procedure of \citet{Ward:2003} to include multiple species as in \citet{Naylor:2010}, we use radiative transfer grids to investigate the physical properties of the CND.  In our preferred model, we correct for a small contribution from the starburst ring for \hco\ \jone\, but assume that all other molecular lines are tracing the CND only.  Radiative transfer analysis cannot substantially constrain the temperature of the CND, but we argue there is evidence for a warm component, so future models should reflect higher gas temperatures.  Though the temperature and density are degenerate, we are able to constrain their product (pressure) well, at Log(P $[\unit{K \ cm^{-3}}]) = 6.60 \pm 0.24$).  The column densities and relative abundances of the molecules, as well as the total mass, are also well constrained.  We also presented models assuming greater contributions to the flux by the starburst ring, and found that these scenarios are difficult to reproduce with well-constrained models. However, one of the models may represent a separate, cooler component that could be distinct from warmer gas.  Should future studies indicate that our preferred model underestimates the contribution of the starburst ring, the results would change systematically as described in \S \ref{sec:sys}.  

In addition to the radiative transfer analysis, we compared the Z-Spec measured line ratios to those predicted by XDR and PDR models.  Our line ratios, measured using a consistent calibration scheme, match extremely well with XDR models at the densities found in our radiative transfer models.  These provide strong independent support of the XDR nature of NGC 1068's CND.  Both the PDR/XDR ratio analysis and the radiative transfer modeling of \hco\ alone indicate that \hco\ may be tracing a different component of gas than the other molecules used in this work (HCN, HNC, CS).  

Future studies of NGC 1068 by the SPIRE instrument on the Herschel Space Observatory will shed further light on the conditions of the CND.  The Fourier Transform Spectrometer should be able to fill out the CO ladder from \jfour\ to \jthirteen\, which will better constrain the temperature of the molecular gas (e.g., \citet{Panuzzo:2010}).  Furthermore, the spectroscopic imaging capability will offer the opportunity to compare the conditions of the CND vs. the starburst ring.  We predict that even some of the higher-J transitions of the high-density tracers from the CND will be measureable with Herschel's HIFI instrument.  ALMA will also be crucial in the further study of NGC 1068, especially to resolve the separate components within the CND.  Higher angular resolution is key to understanding the relation between the gas and the central source, and also important to understanding the clumping and distribution of the gas.

\paragraph{Acknowledgements}
We would like to thank the anonymous referee for a thorough and constructive report, and we would also like to express our gratitude to the staff at the Caltech Submillimeter Observatory.  We acknowledge the following grants and fellowships in support of this
work: NASA SARA grants NAGS-11911 and NAGS-12788, an NSF Career grant
AST-0239270, and a Research Corporation Award RI0928 to J.~Glenn; a
Caltech Millikan fellowship and JPL Director's fellowship to
C.M.~Bradford; an NRAO Jansky fellowship and NSF grant AST-087990 to
J.E.~Aguirre; a NASA GSRP fellowship to L.~Earle; and an NSF GRFP
fellowship to J.~Kamenetzky.


\end{document}